\documentclass[letterpaper,journal]{IEEEtran}

\usepackage{cite}
\usepackage{amsmath,amssymb,amsfonts}
\usepackage{graphicx}
\usepackage{textcomp}
\usepackage{stfloats}
\usepackage{xcolor}
\usepackage{array}
\usepackage{url}
\usepackage{verbatim}
\usepackage{balance}
\def\BibTeX{{\rm B\kern-.05em{\sc i\kern-.025em b}\kern-.08em
		T\kern-.1667em\lower.7ex\hbox{E}\kern-.125emX}}

\usepackage{subfigure}
\usepackage{epstopdf}

\begin{document}

\title{Towards Environment-Aware 6G Network Planning: An Any-to-Any Channel Knowledge Map Approach}

\author{Yilong Chen, Jiaxuan Li, Yuan Guo, Wenhai Lai, Xinhao Li, Kaiming Shen, Guangxu Zhu, \\ Jie Xu, \IEEEmembership{Fellow, IEEE}, Rui Zhang, \IEEEmembership{Fellow, IEEE}, and Zhi-Quan Luo, \IEEEmembership{Fellow, IEEE}
	\thanks{Y. Chen, J. Li, Y. Guo, and J. Xu are with the School of Science and Engineering, Shenzhen Future Network of Intelligence Institute (FNii-Shenzhen), and Guangdong Provincial Key Laboratory of Future Networks of Intelligence, The Chinese University of Hong Kong (Shenzhen), Guangdong 518172, China (e-mail: chenyilong@cuhk.edu.cn; jiaxuanli4@link.cuhk.edu.cn; guoyuan@cuhk.edu.cn; xujie@cuhk.edu.cn).}
	\thanks{W. Lai is with the Queensland Technology International School of Engineering, Dalian University of Technology, Dalian 124221, China (e-mail: wenhailai@dlut.edu.cn).}
	\thanks{X. Li, G. Zhu, and Z.-Q. Luo are with the Shenzhen Research Institute of Big Data, The Chinese University of Hong Kong (Shenzhen), Guangdong 518172, China (e-mail: xinhaoli@link.cuhk.edu.cn, gxzhu@sribd.cn, luozq@cuhk.edu.cn).}
	\thanks{K. Shen is with the School of Science and Engineering, The Chinese University of Hong Kong (Shenzhen), Guangdong 518172, China (e-mail: shenkaiming@cuhk.edu.cn).}
%	\thanks{J. Zhang is with the State Key Laboratory of Networking and Switching Technology, Beijing University of Posts and Telecommunications, Beijing 100876, China (e-mail: jhzhang@bupt.edu.cn).}
	\thanks{R. Zhang is with the Department of Electrical and Computer Engineering, National University of Singapore, Singapore 117583 (e-mail: elezhang@nus.edu.sg).}
	\thanks{J. Xu is the corresponding author.}
}

\maketitle

\begin{abstract}

Planning sixth-generation (6G) networks for reliable coverage across diverse environments is challenging, as it requires channel knowledge between candidate network node locations and potential service regions. This article proposes an environment-aware network planning framework based on any-to-any (X2X) channel knowledge maps (CKMs), which characterize channels between arbitrary transmitter-receiver location pairs. Within this framework, we first construct X2X CKMs from environmental information, sparse channel measurements, or both, with map types and spatial resolutions selected according to the planning task. We then use these maps to predict coverage and optimize network deployment and configuration subject to service requirements and practical constraints, while accounting for subsequent network operation. Finally, field measurements assess channel prediction accuracy and deployment performance, providing feedback for CKM correction and deployment refinement. We discuss implementation methods and tradeoffs across these stages. Case studies illustrate the effectiveness of the framework for indoor intelligent reflecting surface (IRS) deployment and outdoor base station (BS) deployment. Live-network trials further demonstrate coverage gains from IRS deployment and BS reconfiguration, with measurement feedback supporting CKM calibration and configuration refinement.

\end{abstract}

%\begin{IEEEkeywords}
%	Channel knowledge map (CKM), wireless network planning, environment-aware communications.
%\end{IEEEkeywords}

\section{Introduction}

Sixth-generation (6G) wireless networks are expected to support emerging capabilities and applications, including integrated sensing and communication (ISAC), embodied intelligence, and services for the low-altitude economy \cite{ref_Tong2021NextHorizon}. These developments impose increasingly stringent requirements for reliable coverage across diverse environments, including indoor spaces, terrestrial service areas, and low-altitude airspace. Meeting these requirements calls for cooperation among heterogeneous network nodes, including macro, micro, and pico base stations (BSs) and intelligent reflecting surfaces (IRSs). Network planning therefore plays a key role in determining the types, numbers, locations, and configurations of network nodes to satisfy service requirements while minimizing costs. This task is challenging because network coverage depends on the local propagation environment and interactions among deployed nodes. For example, buildings can block direct links, while a suitably located IRS may provide a reflected path around a blockage. Deploying additional BSs can extend coverage, but also increases costs and may introduce interference. Channel knowledge for candidate node locations is therefore essential for evaluating deployment options.

Conventional network planning relies on stochastic channel models and field measurements. Stochastic models are widely used to describe channel statistics, such as path loss, shadow fading, and line-of-sight (LoS) probability, through measurement-calibrated or geometry-based stochastic abstractions \cite{ref_Zhang2020ChannelSurvey}. However, these abstractions may not capture the actual blockage and reflection conditions at candidate deployment locations in a given environment. Field measurements can characterize channels between deployed network nodes and test locations, but these observations do not directly describe links involving nodes at new locations. Thus, obtaining accurate, site-specific channel knowledge for numerous candidate links involving undeployed nodes remains challenging, particularly under practical measurement and computational constraints.

Recently, channel knowledge maps (CKMs) have emerged as a promising approach to environment-aware communications and network deployment. They associate transceiver locations with site-specific channel characteristics, such as channel gains, propagation angles, multipath parameters, or full channel state information (CSI) \cite{ref_CKM1,ref_CKM2}. In particular, large-scale any-to-any (X2X) CKMs are well suited to network planning \cite{ref_CKM2}. Large-scale channel knowledge, such as locally averaged channel gains, LoS states, and the directions of stable propagation paths, characterizes relatively stable spatial propagation conditions and supports long-term planning without requiring predictions of instantaneous small-scale channels. Moreover, X2X CKMs provide channel knowledge between arbitrary transmitter-receiver location pairs within their domains, including candidate locations where nodes have not yet been deployed. This is beneficial over conventional BS-to-any (B2X) CKMs that describe channels between a fixed BS location and different user locations \cite{ref_CKM2}. These properties motivate our focus on large-scale X2X CKMs. Existing studies illustrate the use of X2X CKMs to provide BS-to-IRS and IRS-to-user channel knowledge for IRS deployment \cite{ref_IRS}, as well as ground-to-air channel knowledge for joint air corridor and BS planning \cite{ref_UAV}.

The practical implementation of CKM-enabled network planning faces two main challenges. The first is constructing accurate X2X CKMs from environmental information, limited channel measurements, or both. Environmental information may be inaccurate, and material dielectric properties can be difficult to obtain. Moreover, available measurements are typically collected from existing nodes at a limited number of locations, whereas planning must evaluate links involving transmitters or receivers that have not yet been deployed. Consequently, constructing accurate X2X CKMs for unobserved links without dense measurements remains challenging.

The second challenge is optimizing deployment while accounting for subsequent network operation and channel prediction uncertainty. Deployment options must be evaluated before installation, yet their coverage depends on how transmit power, beamforming, and user association are adapted during operation. These operating decisions, in turn, rely on the deployed nodes and available propagation paths. Furthermore, CKM prediction errors can affect coverage estimates and alter the ranking of candidate deployments. Planning must therefore account for both achievable operating performance and the uncertainty in CKM predictions.

To address these challenges, we propose a CKM-enabled network planning framework comprising three stages:

\begin{itemize}
    \item \emph{Large-scale X2X CKM construction:} We first construct X2X CKMs from environmental information, channel measurements, or both. Construction methods include ray tracing, signal processing, wireless radiance fields, and generative artificial intelligence (AI). We select CKM types and construction methods according to the planning task and adapt the spatial resolution to balance prediction accuracy and computational cost.
    \item \emph{Deployment and configuration optimization:} We then use the constructed CKMs to predict coverage under different deployment schemes and optimize node locations and configurations subject to service requirements and deployment constraints, while accounting for subsequent network operation.
    \item \emph{Field calibration and refinement:} Finally, we use field measurements to assess channel prediction accuracy and deployment performance. The resulting feedback guides CKM calibration and, where necessary, adjustments to node configurations or placement.
\end{itemize}

\noindent We illustrate the proposed framework through simulation case studies of indoor IRS deployment and outdoor BS deployment. Live-network trials further demonstrate coverage improvements over conventional designs through CKM-enabled IRS placement and reconfiguration of an existing BS.

Prior CKM tutorials and surveys cover the CKM concept, construction methods, and applications, including predictive communications at locations yet to be reached \cite{ref_CKM1,ref_CKM2,ref_CKM4}. Related efforts include digital twin channels \cite{ref_DigitalTwinChannel}, which use real-time channel mapping and feedback to support communication operations, and simulated reality of communication networks (SRCON) \cite{ref_SRCON}, which combines propagation, traffic, and user association models to predict network performance. Differently, our framework focuses on CKM-enabled planning for the long-term deployment of active and passive network nodes. It constructs reusable X2X CKMs and uses them to determine node locations and configurations before deployment changes are implemented.

\begin{figure*}[ht]
	\centering
	\includegraphics[width=0.92\linewidth]{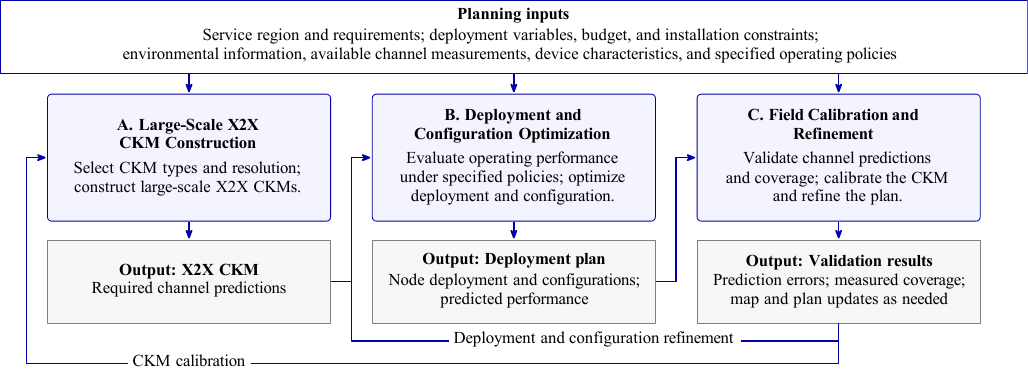}
	\caption{Framework for CKM-enabled network planning. Field measurements support CKM calibration and refinement of deployment and configuration decisions. Post-implementation tests assess deployment performance.}
	\label{fig:framework}
\end{figure*}

\section{CKM-Enabled Network Planning Framework}
\label{sec:framework}

This section presents the CKM-enabled network planning framework as shown in Fig.~\ref{fig:framework}. The framework consists of three stages, including large-scale X2X CKM construction, deployment and configuration optimization, and field calibration and refinement, as detailed in the following subsections.

\subsection{Large-Scale X2X CKM Construction}

In the first stage, we construct large-scale X2X CKMs from environmental information, channel measurements, or both. We first select the CKM types according to the planning task and the types of nodes to be deployed or reconfigured. Channel-gain maps describe the strengths of desired and interfering links, while LoS maps indicate whether an unobstructed direct path is available. These maps are generally needed for coverage analysis. When directional BS antennas or IRSs are considered, dominant path directions or preferred beam indices may also be required \cite{ref_BIM_CAM}.

After selecting the CKM types, we choose construction methods according to the available inputs. Environmental information may include scene geometry and material properties, while channel measurements provide observations at known transceiver locations \cite{ref_CKM2,ref_CKM4}. Representative construction methods include the following.

\begin{itemize}
    \item \emph{Environment-based methods:} These methods simulate wireless propagation using an explicit scene model. Ray tracing is a representative example, which predicts propagation paths and their interactions with objects using scene geometry and material properties \cite{ref_Yun2015RayTracing}. With sufficiently detailed scene information, it can predict channels at candidate locations without direct measurements of those links. Its accuracy generally depends on the fidelity of the reconstructed environment and the propagation effects modeled.
    \item \emph{Data-driven methods:} These methods infer CKMs from channel measurements without requiring an explicit scene model. Conventional approaches, including Kriging interpolation, kernel regression, and matrix completion, recover unobserved values by exploiting spatial correlations in the sampled scene \cite{ref_CKM4}. Their accuracy depends on sampling density and how well the samples capture spatial channel variations around environmental structures. Learning-based methods can use priors learned from datasets spanning multiple scenes to reconstruct CKMs from sparse or low-resolution observations in a new scene. For example, diffusion and other generative models can support map completion or resolution enhancement \cite{ref_CKMDiff}. Their performance in a new scene depends on how well the training data represent its propagation patterns.
    \item \emph{Joint environment- and data-based methods:} These methods combine environmental information and channel measurements for CKM construction. Representative approaches include physics-informed learning, which integrates propagation models with measurements to fit a scene representation. For example, wireless radiance fields use neural fields or explicit Gaussian primitives to represent propagation properties, with parameters learned from scene-specific measurements. Once fitted, these representations can predict channels at unmeasured locations within the same scene. X2X CKM construction further requires support for varying transmitter and receiver locations. Bidirectional representations like bidirectional wireless Gaussian splatting provide one way to meet this requirement \cite{ref_BiWGS}.
    Virtual-scatterer-based channel modeling is another promising approach by representing a physical environment with a set of virtual scatterers, and by optimizing their number, positions, and propagation parameters with channel measurements at known locations, the CKM at unknown locations can be reconstructed efficiently \cite{sun2026channel}.
\end{itemize}

Practical CKM construction involves a tradeoff among prediction accuracy, spatial resolution, and construction cost. This is particularly important over large deployment regions, where detailed propagation modeling and dense sampling can be expensive. Two approaches can reduce the construction effort. First, we can adapt the map resolution to the planning task and the requirements of different service regions. For example, we can construct a coarse map over the entire region and refine it where additional detail is needed to distinguish deployment options. Next, we can separate environmental propagation information in the CKM representation from device characteristics. The same propagation representation can then be combined with the corresponding device properties to predict coverage for different node types and deployment schemes.

\subsection{Deployment and Configuration Optimization}

In the second stage, we use the constructed X2X CKMs to predict network performance and optimize network deployment and node configurations. Here, deployment decisions specify the types, numbers, locations, and orientations of network nodes, while node configurations include adjustable settings such as BS transmit powers and beams or IRS reflection coefficients. Some configurations are selected during planning, whereas others are adapted during operation according to specified policies. The planning task may involve deploying a new network, adding nodes to an existing network, or reconfiguring installed nodes. These decisions must satisfy practical constraints on permitted mounting regions, allowable node numbers and types, placement and orientation, and feasible power, antenna, and reflection settings.

Network planning faces several coupled challenges. First, the design must balance potentially competing objectives, such as network coverage, achievable rate, operating energy consumption, and deployment cost. Next, jointly selecting node types, numbers, locations, orientations, and configurations creates a large search space containing discrete and continuous variables. Interference and cascaded propagation further introduce nonconvex relationships. Furthermore, deployment performance depends on decisions made at a shorter timescale, including power control, beamforming, resource allocation, and user association. This dependence leads to a nested problem: an outer optimization selects deployment decisions, while an inner evaluation or optimization determines their operating performance under the assumed policies. In addition, large-scale CKMs support long-term decisions, but adaptive BS transmission and IRS reflection settings additionally depend on instantaneous channel information that is unknown during planning.

To tackle these challenges, we select optimization methods according to the variable structure and the available performance evaluator. First, large-neighborhood search and genetic algorithms can explore discrete deployment choices, with locations, heights, orientations, and configuration options discretized when appropriate. Next, continuous locations and configurations can be refined using gradient-based methods when the CKM model and the full performance evaluation provide usable derivatives. Linearization and convex approximation can be used to simplify structured subproblems, while valid relaxations can yield tractable approximations or performance bounds. Furthermore, when the evaluator exposes only performance values, derivative-free methods can search the resulting black-box objective. Learning-based methods can also use solved instances or evaluation feedback to guide candidate selection and node configuration. In practice, the optimization and variable search must balance deployment performance against the computational and memory costs of CKM queries and network performance evaluation. Precomputing frequently reused channel values and caching query results can reduce repeated computation, while on-demand queries reduce storage requirements.

Uncertainty in CKM predictions further affects performance estimates and the feasibility of candidate deployments. These prediction errors should be distinguished from the small-scale fading accounted for when evaluating network operation. When suitable error estimates are available, candidate deployments can be evaluated under plausible CKM prediction errors, and robust or probabilistic formulations can incorporate this uncertainty into planning objectives and constraints. Such formulations require appropriate uncertainty sets or calibrated probability models. Additional measurements collected during the subsequent field calibration and refinement stage can support local CKM refinement and updates to these uncertainty models, allowing deployment decisions to be reassessed using improved channel knowledge.

\subsection{Field Calibration and Refinement}

In the third stage, we use field measurements to validate channel predictions and refine deployment decisions. Measurement feedback supports CKM construction and deployment optimization, while tests conducted after implementation assess whether the deployed network meets service requirements.

To calibrate the CKM, we examine prediction errors that may arise from limited or noisy measurements, inaccurate scene geometry or material properties, or environmental changes. Depending on the error patterns and construction method, we can correct map values by fitting a spatial correction to measured discrepancies or update the underlying model through parameter calibration or predictor refitting. For conventional ray tracing, calibration may involve refining scene geometry and the material properties assigned to environmental objects. For differentiable ray tracing, model parameters can be optimized using channel measurements; wireless radiance fields, virtual scatterers, and other learned predictors can be refitted or fine-tuned with the new data.

For deployment validation and refinement, we compare measured and predicted coverage and assess whether service requirements are met at the tested locations. When an existing network is modified, measurements taken at the same locations before and after implementation quantify coverage changes under comparable operating conditions. We then reoptimize the deployment using the calibrated CKM, incorporating robust or probabilistic service constraints when measurements support a suitable uncertainty set or probability model for the residual errors. If service requirements are not met, we first adjust node configurations, such as BS transmit powers and beams or IRS reflection coefficients. Feasible local placement changes are considered when configuration adjustments are insufficient. Further measurements at additional locations can assess the corrected CKM and determine whether the deployment adjustments improve coverage.

\section{CKM-Enabled Network Planning Simulator}
\label{sec:simulator}

%\begin{figure}[t]
%	\centering
%	\includegraphics[width=0.95\linewidth]{figure_agent.pdf}
%	\caption{CKM-enabled planning workflows for indoor IRS and outdoor BS deployment.}
%	\label{fig:case-studies}
%\end{figure}

This section presents a CKM-enabled network planning simulator that implements the CKM construction and deployment optimization stages of the proposed framework.\footnote{The source code is available at \url{https://github.com/chenyilong-cuhksz/ckm-planning-agent}.} We use the simulator for two representative case studies: IRS deployment and reflection configuration for indoor coverage enhancement, and BS deployment for outdoor coverage enhancement.

The construction and optimization workflow proceeds as follows. First, we obtain the scene geometry, which can be reconstructed from multiview scans for indoor environments or obtained from OpenStreetMap\footnote{OpenStreetMap: \url{https://www.openstreetmap.org/}.} for outdoor environments. We combine this geometry with material information to build radio scene models. Next, we identify feasible installation regions and target service regions according to the geometric constraints of each scene. We also specify the candidate node budget and the locations of any existing nodes. We then sample candidate node sites within the installation regions and define receiver grids within the service regions for performance evaluation. Using Sionna RT\footnote{Sionna RT: \url{https://nvlabs.github.io/sionna/rt/}.}, we compute propagation paths between candidate node locations and receiver grid points, from which we construct CKMs containing locally averaged channel gain predictions for these links. Finally, we use the constructed CKMs to optimize node locations and configurations to improve coverage over the service regions. The following subsections present the two case studies in detail.

\begin{figure}[t]
	\centering
	\includegraphics[width=1.0\linewidth]{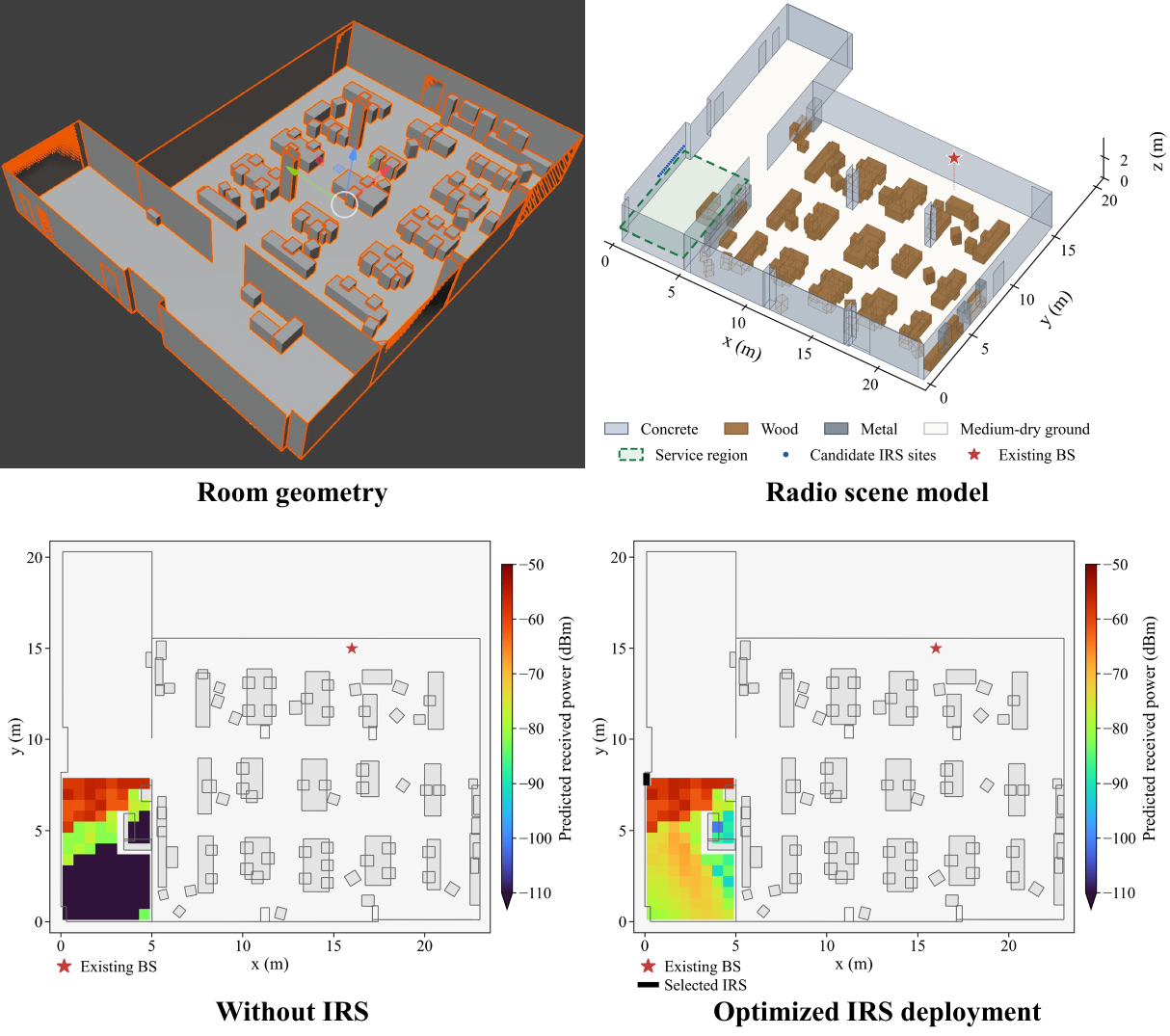}
	\caption{Scene modeling and simulation results for CKM-enabled indoor IRS deployment. The received-power results compare the cases without IRS and with the optimized IRS deployment.}
	\label{fig:case-studies1}
\end{figure}

\begin{figure}[t]
	\centering
	\includegraphics[width=1.0\linewidth]{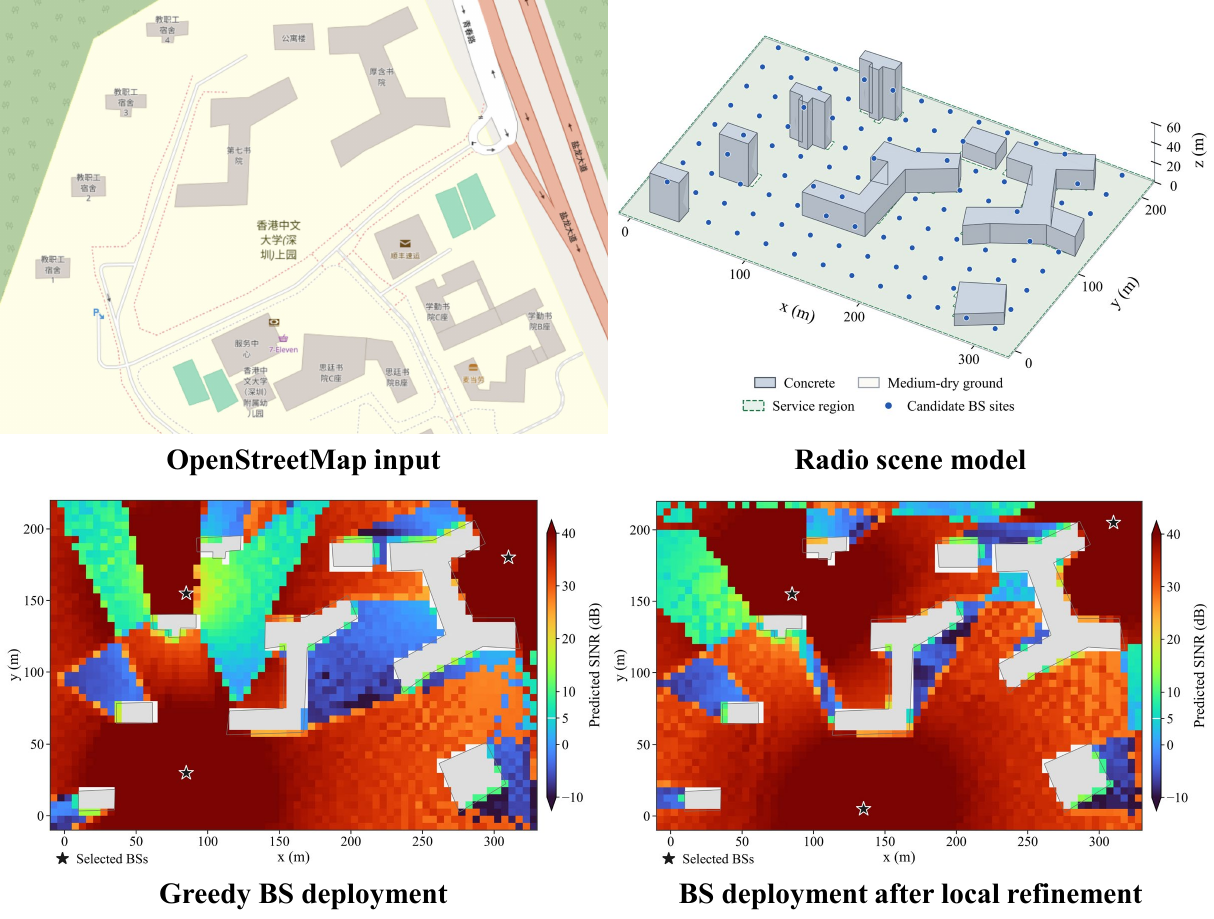}
	\caption{Scene modeling and simulation results for CKM-enabled outdoor BS deployment. The SINR results compare the greedy layout with the layout after local site refinement.}
	\label{fig:case-studies2}
\end{figure}

\subsection{Indoor IRS Deployment}

We first consider an indoor IRS deployment scenario based on a real office at CUHK-Shenzhen, as shown in Fig.~\ref{fig:case-studies1}. A single-antenna BS serves the room, where a wall blocks signal propagation and creates a coverage hole behind it. The BS location, service region, and permitted mounting wall are specified before planning.
The objective is to deploy one IRS on this wall and configure its reflection to maximize the average received power over the service region. The BS transmit power is 10 dBm, and the IRS consists of 16 $\times$ 16 reflecting elements with four-level phase control.

Following the simulator workflow, we reconstruct the room geometry from multiview scans and assign material properties to the scene objects. 
We sample 16 candidate IRS sites on the permitted wall and discretize the service region into 97 square grids, each with a side length of 0.6 m. We then construct CKMs for the BS-to-receiver, BS-to-IRS, and IRS-to-receiver links associated with these locations. 
To account for the coupling between IRS placement and reflection configuration, we use a nested optimization procedure that evaluates each candidate site under a specified reflection design policy \cite{ref_IRS}. For each candidate location, the inner step generates quantized reflection patterns based on channel phase alignment, calculates the corresponding reflecting response, and evaluates the resulting channel gain over the service region. The outer step selects the IRS location according to the resulting average received power.

Fig.~\ref{fig:case-studies1} shows the simulated received-power results.
Without the IRS, the average received power over the service region is -90.85 dBm. The selected IRS deployment increases this value to -72.03 dBm, compared with an average of -74.42 dBm across the 16 candidate-site deployments. These simulation results illustrate the benefit of using IRS-related channel knowledge to select a deployment site under the mounting constraints.

\subsection{Outdoor BS Deployment}

We next consider deploying three BSs to provide outdoor coverage on the CUHK-Shenzhen campus, as shown in Fig.~\ref{fig:case-studies2}. Each BS is equipped with a single antenna and transmits at 10 dBm. The objective is to maximize the signal-to-interference-plus-noise ratio (SINR)-based coverage ratio over the service region, defined as the fraction of receiver grids with an SINR of at least 5 dB. We associate each grid with the BS that provides the highest received power and treat the other deployed BSs as cochannel interferers.

We obtain the outdoor scene geometry from OpenStreetMap and assign material properties to the buildings. Geometric and clearance checks identify 98 feasible candidate sites within the installation region. The service region is discretized into 2,741 square grids, each with a side length of 5 m. We construct a CKM containing channel gains between candidate BS sites and receiver grids to evaluate different three-BS layouts. We then optimize the deployment through greedy site selection followed by local site refinement.

As shown in Fig.~\ref{fig:case-studies2}, the resulting deployment achieves a coverage ratio of 88.1\%, compared with 81.2\% for greedy selection alone. Uniform random selection of three BS sites from the same 98 candidates yields an average coverage ratio of 55.9\%, evaluated over all possible three-site combinations. These simulation results illustrate how CKM-enabled BS site selection can improve coverage by accounting for both desired signal strength and inter-BS interference.

\section{Field-Test Results}

This section presents two live-network trials of CKM-enabled planning for indoor IRS deployment and outdoor BS reconfiguration, respectively. We compare the measured signal coverage with conventional designs and examine how measurement feedback supports CKM calibration and node configuration.

\subsection{Indoor Coverage Enhancement via IRS Deployment}

\begin{figure}[t]
	\centering
	\includegraphics[width=0.95\linewidth]{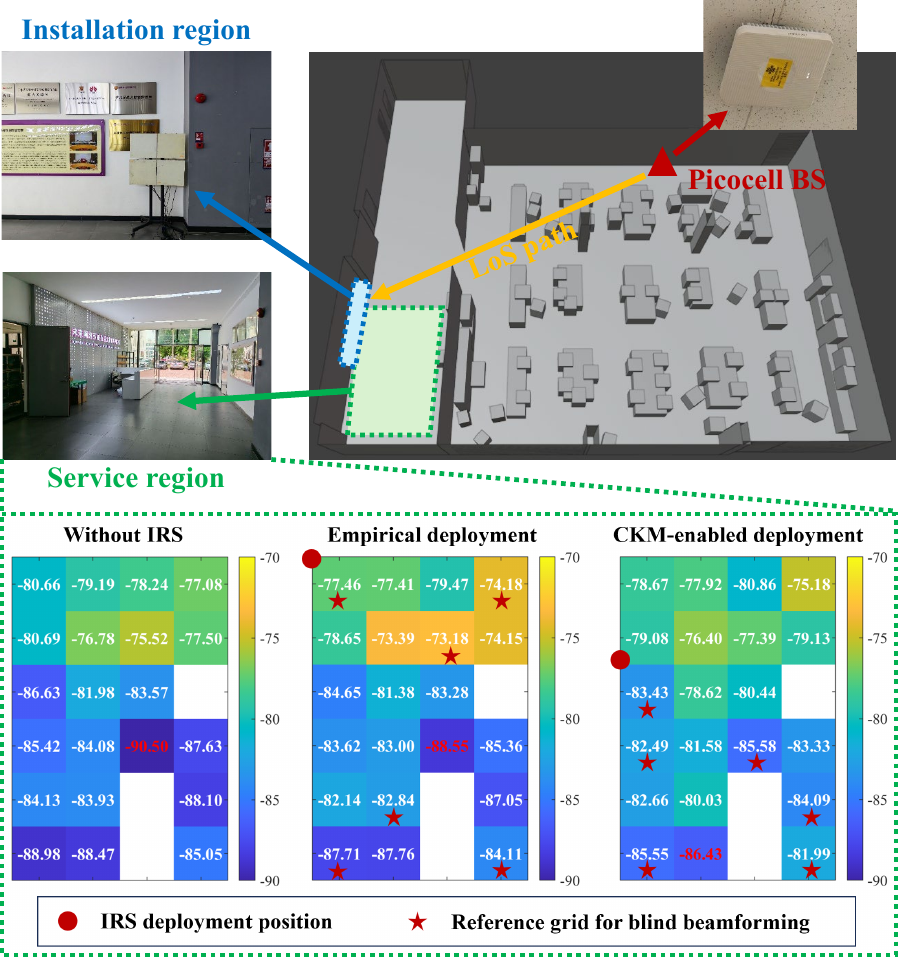}
	\caption{Indoor live-network trial of CKM-enabled IRS deployment: test scene and measured received power. The results compare no IRS, empirical IRS deployment, and CKM-enabled IRS deployment.}
	\label{fig:indoor}
\end{figure}

%\begin{figure}[t]
%	\centering
%	\includegraphics[width=1.0\linewidth]{figure_indoor2.pdf}
%	\caption{Measured received power in the service region without an IRS, with empirical IRS deployment, and with CKM-enabled IRS deployment.}
%	\label{fig:indoor2}
%\end{figure}

As shown in Fig.~\ref{fig:indoor}, the indoor trial considers the same indoor scene as in Section~\ref{sec:simulator}, where a 3.5 GHz picocell BS is mounted on the ceiling of a hall. The service region is an adjacent stairwell, where blockage results in weak coverage. The objective is to improve coverage by deploying an IRS within the permitted installation region. For evaluation, we divide the service region into 21 grids, each with a side length of 1.2 m, and use a smartphone to measure the average received power within each grid.

We construct the CKM following the procedure in Section~\ref{sec:simulator}. In particular, the large-scale CKM is used to select the IRS installation site and reference grids in weak-coverage regions, as obtaining small-scale CSI for all candidate IRS-receiver links would require extensive field measurements. Once the IRS is installed, we configure its reflections using blind beamforming \cite{xu2024blind}, which relies on received-power feedback from the selected reference grids without requiring explicit cascaded CSI. This feedback allows the reflection settings to be adapted to the propagation conditions at the installed site.

We compare the CKM-enabled scheme with two benchmarks: no IRS and empirical IRS deployment. In the empirical scheme, both the installation site and the reference grids for blind beamforming are selected manually based on scene geometry. The comparison therefore evaluates the combined benefit of using CKM information for these two planning decisions.

Fig.~\ref{fig:indoor} shows the measured received-power results for the three schemes. Relative to the case without an IRS, empirical deployment improves the average and minimum received power across the weak-coverage regions by 1.31 dB and 1.95 dB, respectively. CKM-enabled deployment increases these gains to 3.25 dB and 4.07 dB, corresponding to additional gains of 1.94 dB in average received power and 2.12 dB in minimum received power over the empirical scheme. These results demonstrate the benefit of CKM-enabled site and reference grid selection, followed by IRS configuration using received-power feedback, for coverage enhancement in the tested region.

\subsection{Outdoor Coverage Enhancement via BS Reconfiguration}
\label{sec:live-network-bs-tuning}

\begin{figure}[t]
	\centering
	\includegraphics[width=1.0\linewidth]{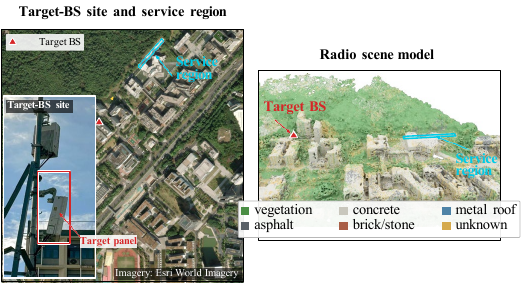}
	\caption{Outdoor live-network trial of CKM-enabled BS reconfiguration: target-BS site, service region, and radio scene model.}
	\label{fig:outdoor-scene}
\end{figure}

\begin{figure}[t]
	\centering
	\includegraphics[width=0.95\linewidth]{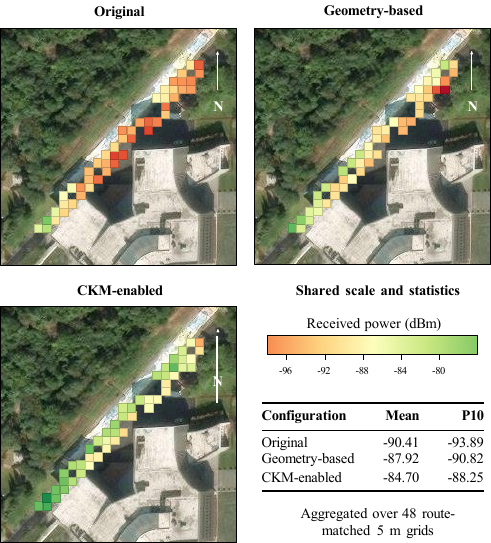}
	\caption{Measured received power and statistics for outdoor BS reconfiguration under the original, geometry-based, and CKM-enabled configurations. The mean and P10 are calculated over the same 48 measurement grids.}
	\label{fig:outdoor-results}
\end{figure}

The outdoor trial considers reconfiguring an existing BS to improve coverage over the service region shown in Fig.~\ref{fig:outdoor-scene}. Note that deploying additional BSs in an operational network involves installation costs and operator approval, and as a result, we focus on reconfiguration that reuses the existing infrastructure.

We evaluate signal coverage using synchronization signal-based reference signal received power (SS-RSRP). The target-BS site, antenna hardware, and mechanical orientation remain fixed. The software-configurable settings include vendor-supported coverage modes that specify the beamwidth and allowable steering ranges, electronic azimuth offset and tilt that control the beam direction relative to the installed panel, and synchronization signal block (SSB) power bias that adjusts the reference-signal level.

The reconfiguration follows a measurement-assisted CKM workflow. We first import registered scene geometry and material information into Sionna RT to construct an initial CKM that captures site-specific propagation and received-power sensitivity to beam direction. Measurements collected under the original configuration then calibrate the absolute received-power predictions through an additive spatial residual correction. For each candidate configuration, we combine the calibrated map for the original configuration with the spatial change predicted by Sionna RT for the beam adjustment and the corresponding SSB power offset. We assess feasible configurations using their predicted average received power and received power in weak-coverage regions, while accounting for channel prediction errors and the magnitude of parameter changes.

We test three configurations: the original setting, a geometry-based setting, and our proposed CKM-enabled setting. Here, the geometry-based setting is determined according to the target direction, distance, height, and nominal beam footprint. After applying each configuration, we aggregate measurements collected along the same route into 48 common grids, each with a side length of 5 m. We compare the mean received power and its spatial tenth percentile (P10) across these grids, with P10 characterizing the weaker end of the measured signal distribution.

Fig.~\ref{fig:outdoor-results} presents the measured received-power results and statistics. CKM-enabled reconfiguration improves the mean received power by 5.71 dB over the original setting and by 3.22 dB over the geometry-based setting. The corresponding P10 gains are 5.64 dB and 2.57 dB, respectively. Moreover, received power improves in all 48 matched grids relative to the original configuration. These gains demonstrate the benefit of the selected configuration, including its beam and power settings, for signal coverage along the measured route. The trial also illustrates how initial measurements support CKM calibration, while measurements after reconfiguration validate the resulting coverage and provide a basis for further refinement.

\section{Conclusion}

In this article, we proposed a CKM-enabled network planning framework comprising three stages: large-scale X2X CKM construction, deployment and configuration optimization, and field calibration and refinement. Simulation case studies illustrated its application to indoor IRS deployment and outdoor BS deployment, while live-network trials demonstrated coverage gains from IRS deployment and BS reconfiguration. Measurement feedback supported CKM calibration and node configuration. Efficient X2X CKM construction and robust deployment optimization remain research directions for extending the framework to larger regions and more diverse network node types.

\bibliographystyle{IEEEtran}
\bibliography{ref2}

\end{document}